\documentclass[%
 reprint,
 superscriptaddress,
 amsmath,amssymb,
 aps,
]{revtex4-2}
\usepackage{xcolor}
\usepackage{graphicx}
\usepackage{dcolumn}
\usepackage{bm}
\usepackage{soul}

\begin{document}
\preprint{APS/123-QED}

\title{Asymmetric Faraday effect caused by a break of spatial symmetry}

\author{D.O. Ignatyeva}
\email[]{ignatyeva@physics.msu.ru}
\affiliation{Photonic and Quantum technologies school, Faculty of Physics, Lomonosov Moscow State University, Leninskie gori, 119991 Moscow, Russia}
\affiliation{Russian Quantum Center, 121205 Moscow, Russia}
\affiliation{Institute of Physics and Technology, V.I. Vernadsky Crimean Federal University, 295007 Simferopol, Crimea}

\author{T.V. Mikhailova}
\affiliation{Institute of Physics and Technology, V.I. Vernadsky Crimean Federal University, 295007 Simferopol, Crimea}

\author{P.O. Kapralov}
\affiliation{Russian Quantum Center, 121205 Moscow, Russia}

\author{S.D. Lyashko}
\affiliation{Institute of Physics and Technology, V.I. Vernadsky Crimean Federal University, 295007 Simferopol, Crimea}

\author{V.N. Berzhansky}
\affiliation{Institute of Physics and Technology, V.I. Vernadsky Crimean Federal University, 295007 Simferopol, Crimea}

\author{V.I. Belotelov}
\affiliation{Institute of Physics and Technology, V.I. Vernadsky Crimean Federal University, 295007 Simferopol, Crimea}
\affiliation{Photonic and Quantum technologies school, Faculty of Physics, Lomonosov Moscow State University, Leninskie gori, 119991 Moscow, Russia}
\affiliation{Russian Quantum Center, 121205 Moscow, Russia}

\date{\today}

\begin{abstract}
It is widely known that the magneto-optical Faraday effect is linear in magnetization, and therefore the Faraday angles for the states with opposite magnetizations are of opposite sign but equal in modulus. Here we demonstrate that under certain spatial symmetry-breaking conditions, an asymmetric Faraday effect (AFE) arises, meaning that the Faraday angles for opposite magnetic states differ not only in sign but in absolute value as well. Experimental investigations of AFE are performed in a one-dimensional all-garnet magnetophotonic crystal, where AFE appears in the vicinity of the cavity resonance for an oblique incidence of light with an inclined light polarization plane. The magnitude of the observed asymmetry between Faraday rotations for the two opposite magnetizations is very large and reaches 30$\%$ of the absolute value of the Faraday effect. We confirm the generality of the suggested effect by the numerical analysis
of several different configurations in which AFE arises. The discovered AFE is of prime importance for nanoscale magnonics and optomagnetism.
\end{abstract}

\maketitle


\section{Introduction}
The first phenomenon that demonstrated interaction between magnetism and optics and established a basis for modern magnetophotonics is the Faraday effect, which is the transformation of the polarization of light passing through a magnetized material. It was discovered by Michael Faraday in 1845~\cite{faraday1846magnetization}. The Faraday effect is characterized by the Faraday rotation angle of the polarization plane of light transmitted through a material along its magnetization. The internal mechanism of the effect is magnetic circular birefringence, i.e., the magnetization-induced difference in the phase velocities of two circular polarizations with opposite helicities \cite{zvezdin1997modern}. The Faraday effect is known to be odd in magnetization and thus has a nonreciprocal character~\cite{krumme1984measurement,dionne1994molecular,deb2012magneto,caloz2018electromagnetic}. Initially, Faraday's discovery had a purely fundamental significance, but a lot of applications of the magneto-optics for data storage, fiber-optic communication lines (isolators, modulators, deflectors)~\cite{karki2019toward, ho2018switchable,smigaj2010magneto}, sensors~\cite{maccaferri2015ultrasensitive,ignatyeva2016high, borovkova2020high} and magnetometry~\cite{ignatyeva2021vector,knyazev2018magnetoplasmonic} have been suggested in the following decades~\cite{inoue2013magnetophotonics}. For further progress in this direction, it is necessary to find some ways for enhancement and advanced control of the Faraday effect.

Fabrication of nanostructures is the most common solution that has been widely used over some decades to increase the magneto-optical (MO) effects. It is possible to design a system which exhibits an enhanced MO response by concentrating the electromagnetic field of light inside magneto-optically active components in the magnetophotonic crystals (MPCs) with localized states~\cite{inoue2006magnetophotonic,zhdanov2006enhancement,lyubchanskii2003magnetic,maccaferri2020nanoscale,merzlikin2007controllable,mikhailova2018optimization,yu2020nonreciprocal}, guided modes~\cite{voronov2020magneto, chernov2020all,royer2020enhancement,bsawmaii2020longitudinal}, in magnetoplasmonic structures~\cite{belotelov2006magnetooptics,lopez2020enhanced,kalish2018magnetoplasmonic,maccaferri2015resonant,knyazev2018magnetoplasmonic,armelles2013magnetoplasmonics,lodewijks2014magnetoplasmonic,khramova2019resonances} and metasurfaces~\cite{ignatyeva2020all,qin2020switching,bi2021magnetically,yang2022observation,xia2022circular}. Enhancement of the Faraday rotation most often occurs due to an increase in purely magneto-optical contribution (multi-pass mode and localization of light). Additionally, such structures may include the pure optical effects of anisotropy or reducing the reflectivity to enhance the conversion of polarization state (magnetoplasmonic structures~\cite{fujikawa2008contribution,baryshev2013peculiarities,fan2019magneto} or multilayer structures at oblique incidence~\cite{baek2011multiple,grishin2019waveguiding}). The combination of pure magneto-optical and optical mechanisms is responsible for various types of the Faraday effect enhancement~\cite{maccaferri2020nanoscale,armelles2013magnetoplasmonics,grishin2019waveguiding,tomilin2020vertical,vasiliev2006effect,ignatyeva2020bound,doi:10.1063/1.2362987}.

Oblique incidence of light onto magnetic films and structures brings some peculiarities of the MO effects~\cite{schafer2021analyzer} which are related to a lower reflectance of p-polarized light with respect to s-polarization. Thus, the odd MO intensity effect appears in the Faraday configuration~\cite{krinchik1978magneto}. In MPCs, the Fresnel coefficients at oblique incidence set different interference conditions for s- and p-polarized light. Thus, the same MPC has a higher optical Q-factor for s-polarized light and lower Q-factor for p-polarized light~\cite{grishin2019waveguiding,vasiliev2006effect,ignatyeva2020bound,doi:10.1063/1.2362987,wu2021quasi}. An s-polarized wave demonstrates a higher localization of electromagnetic field inside the MO active cavity and layers, and, therefore, higher values of the Faraday rotation~\cite{grishin2019waveguiding}. Furthermore, for MPC with negligible absorption, a wave converted into s-polarized state due to the Faraday effect can be trapped inside the MO active cavity at the Brewster's angle~\cite{ignatyeva2020bound}. 

In all of the aforementioned studies, the Faraday rotation was odd in magnetization, i.e., reversing the direction of the external magnetic field changed its sign but not its magnitude (see Supplementary S1). In this work, we discover that an additional spatial symmetry break might lead to an even magnetization contribution to the magneto-optical Faraday rotation and, therefore, the emergence of the asymmetric Faraday effect (AFE), which changes not only its sign but also its value if the magnetic field is reversed. 
We theoretically analyze the general conditions of spatial symmetry breaking under which the Faraday effect might not be odd. AFE is observed experimentally in a microcavity MPC with obliquely incident light with incline polarization. The difference between the magnitudes of the Faraday rotations of the two oppositely magnetized MPCs reaches $30\%$. We confirm the generality of the suggested effect by the numerical analysis of several different configurations of the other structures in which AFE is present.

\section{Spatial symmetry and the oddness of the Faraday effect}

\begin{figure}[ht]
\centering
\includegraphics[width=0.9\columnwidth]{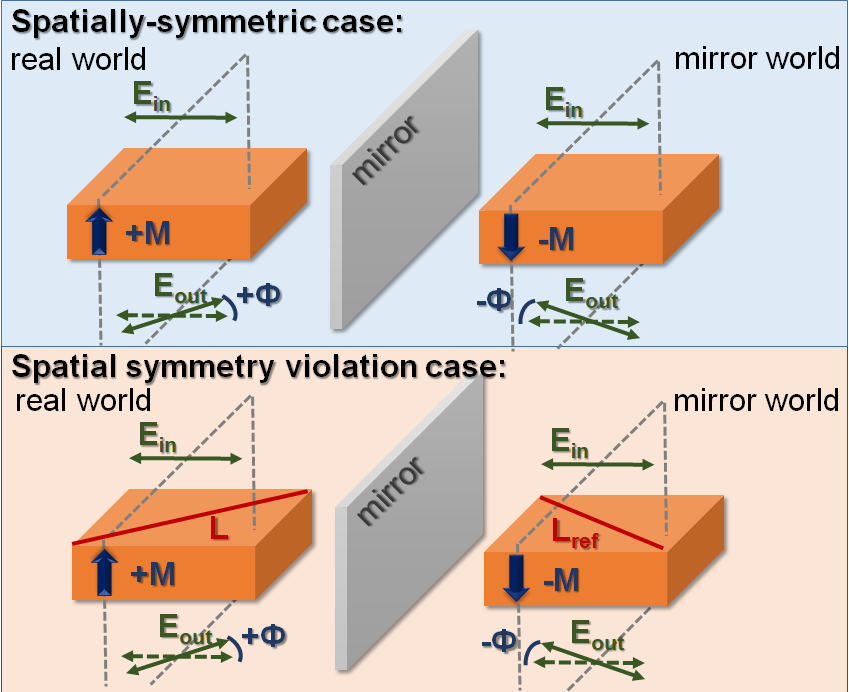}
\caption{Schematic representation of spatial symmetry origin of the oddness of the Faraday effect and its violation in the case of symmetry breaking by a director $\mathbf{L}$.}
\label{fig: sketch of symmetry}
\end{figure}

Let's discuss how the spatial symmetry causes the oddness of the Faraday rotation in the usual case and how it is modified if the spatial symmetry is broken. Figure~\ref{fig: sketch of symmetry} illustrates the discussed phenomenon from the point of view of what an observer might see in the real world and looking into a mirror. The mirror is placed in the plane of the magnetization $\mathbf{M}$ perpendicular to the polarization $\mathbf{E}_\mathrm{in}$ of an incident light, see Fig.~\ref{fig: sketch of symmetry}. An observer sees in the mirror an oppositely magnetized medium with $-\mathbf{M}$, as $\mathbf{M}$ is an axial vector. At the same time, the director (a director is a quasi-vector with two equivalent opposite directions) $\mathbf{E}$ describing the light polarization is transformed by the usual reflection rules in the mirror. Thus, the angle between the magneto-optically rotated $\mathbf{E}_\mathrm{out}$ and the incident $\mathbf{E}_\mathrm{in}$ observed in the mirror is opposite to the angle between the real $\mathbf{E}_\mathrm{out}$ and $\mathbf{E}_\mathrm{in}$. It means that the magneto-optical polarization rotation caused by passing through a magnetized medium that is observed in the mirror world with $-\mathbf{M}$ magnetization is equal in magnitude and opposite in sign to the one observed in the real world (Fig.~\ref{fig: sketch of symmetry}): $-\Phi(\mathbf{M})=\Phi(-\mathbf{M})$ and, therefore, the magneto-optical rotation is odd in magnetization.

The situation changes if the spatial symmetry of the considered configuration (including both the light and the structure) is broken with respect to the reflection in the mirror placed along $\mathbf{M}$ by a presence of the preferred direction described by a director $\mathbf{L}$ (see Fig.~\ref{fig: sketch of symmetry}, bottom panel). Actually, such a preferred direction could arise due to the presence of some vector describing the light or the structure in the considered configuration, too. It might be considered a partial case of the symmetry break described by a director; therefore, in the following, we will consider a more general case of a director. Since a director is transformed by the usual reflection rules, the angle between the director $\mathbf{L}^\mathrm{ref}$ and polarization $\mathbf{E}_\mathrm{in}$ in a mirror is opposite to the one observed in the real world. The exceptions are the cases where $\mathbf{L}$ is perpendicular or parallel to $\mathbf{E}_\mathrm{in}$. Thus, the whole configuration in the mirror is different from what is in the real world not only due to the magnetization reversal, but also due to the different $\mathbf{L}^\mathrm{ref}$ direction.
By repeating the considerations similar to the ones made above, one concludes that $\Phi(-\mathbf{M}, \mathbf{L}^\mathrm{ref})=-\Phi(\mathbf{M},\mathbf{L})$, so the absolute values of $\Phi$ that are equal to each other refer to different structures: $|\Phi(\mathbf{M},\mathbf{L}^\mathrm{ref})|=|\Phi(\mathbf{M},\mathbf{L})|$. At the same time, there are no limitations on the values of $\Phi$ for the structures with the same $\mathbf{L}$;  $|\Phi(-\mathbf{M}, \mathbf{L})|$ and $|\Phi(\mathbf{M},\mathbf{L})|$ may differ from each other. Consequently, we can conclude that if $\mathbf{L}$ is oblique with respect to the polarization $\mathbf{E}$, then the asymmetric Faraday effect (AFE) is allowed: $|\Phi(\mathbf{M},\mathbf{L})|\neq|\Phi(-\mathbf{M}), \mathbf{L}|$. However, if $\mathbf{L}$ is parallel or perpendicular to $\mathbf{E}$, then the AFE is prohibited: $|\Phi(\mathbf{M})|=|\Phi(-\mathbf{M})|$. 

\begin{figure}[htb]
\centering
(a)~~~~~~~~~~~~~~~~~~~~~~~~~~~~~~~~~~~~~~~~~~~~(b)\\
\includegraphics[width=0.99\linewidth]{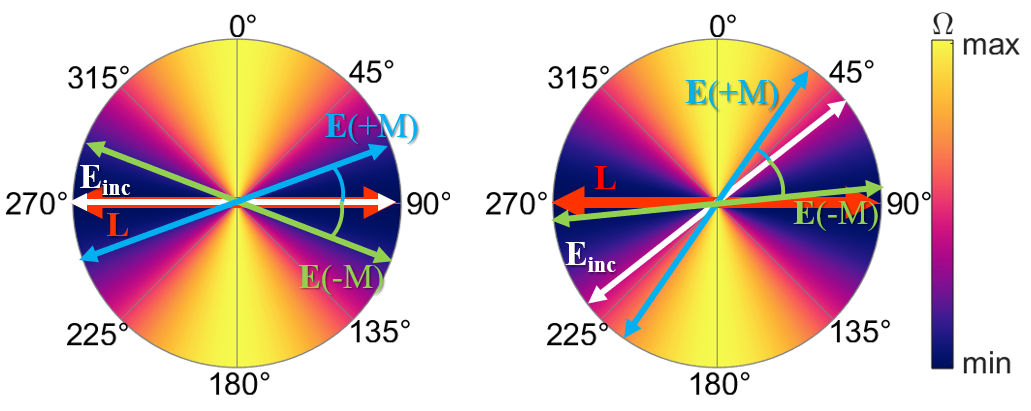}
\caption{Schematical depiction of an asymmetric Faraday effect arising as the result of non-equivalence of the counter- and clockwise rotation directions caused by the presence of the director L along the $x$-axis and consequent polarization dependence of some optical characteristic $\Omega$ (transmittance, absorption, etc.). (a) Case of $\mathbf{E}$ parallel to L and equivalence of clock and counter-clockwise rotations. (b) Case of $\mathbf{E}$ tilted with respect to L which cause non-equivalence of clock and counter-clockwise rotations. }
\label{fig: SI  clock rot}
\end{figure}

\begin{figure*}[ht]
\centering
(a)~~~~~~~~~~~~~~~~~~~~~~~~~~~~~~~~~~~~~~~~~~~~~~~~~~(b)~~~~~~~~~~~~~~~~~~~~~~~~~~~~~~~~~~~~~~~~~~~~~~~~~~(c)\\
\includegraphics[width=0.3\linewidth]{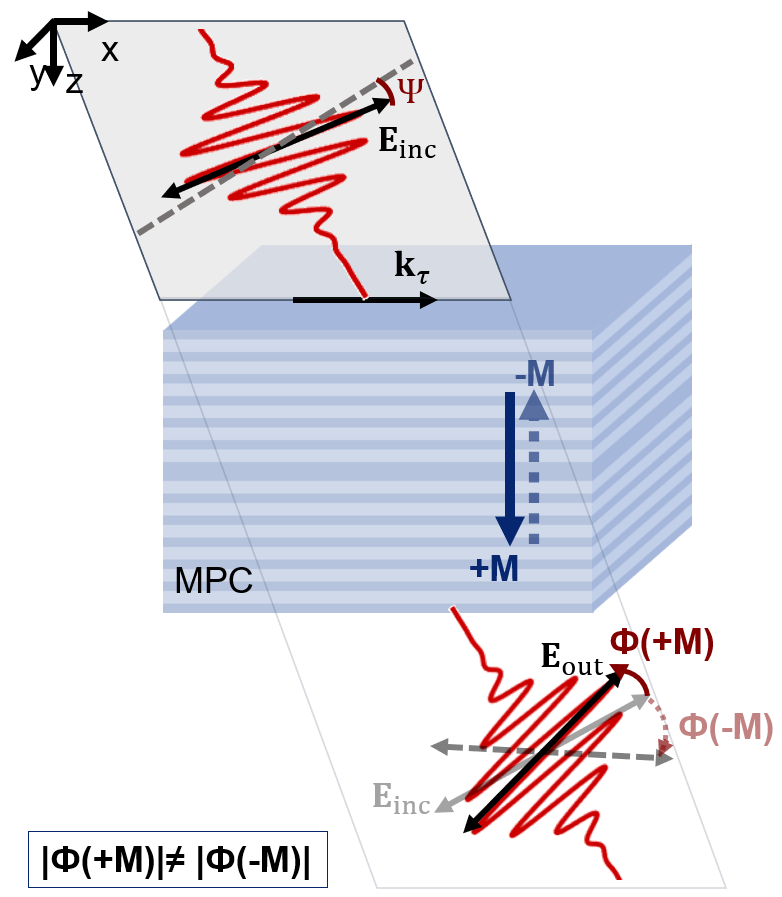}
\includegraphics[width=0.33\linewidth]{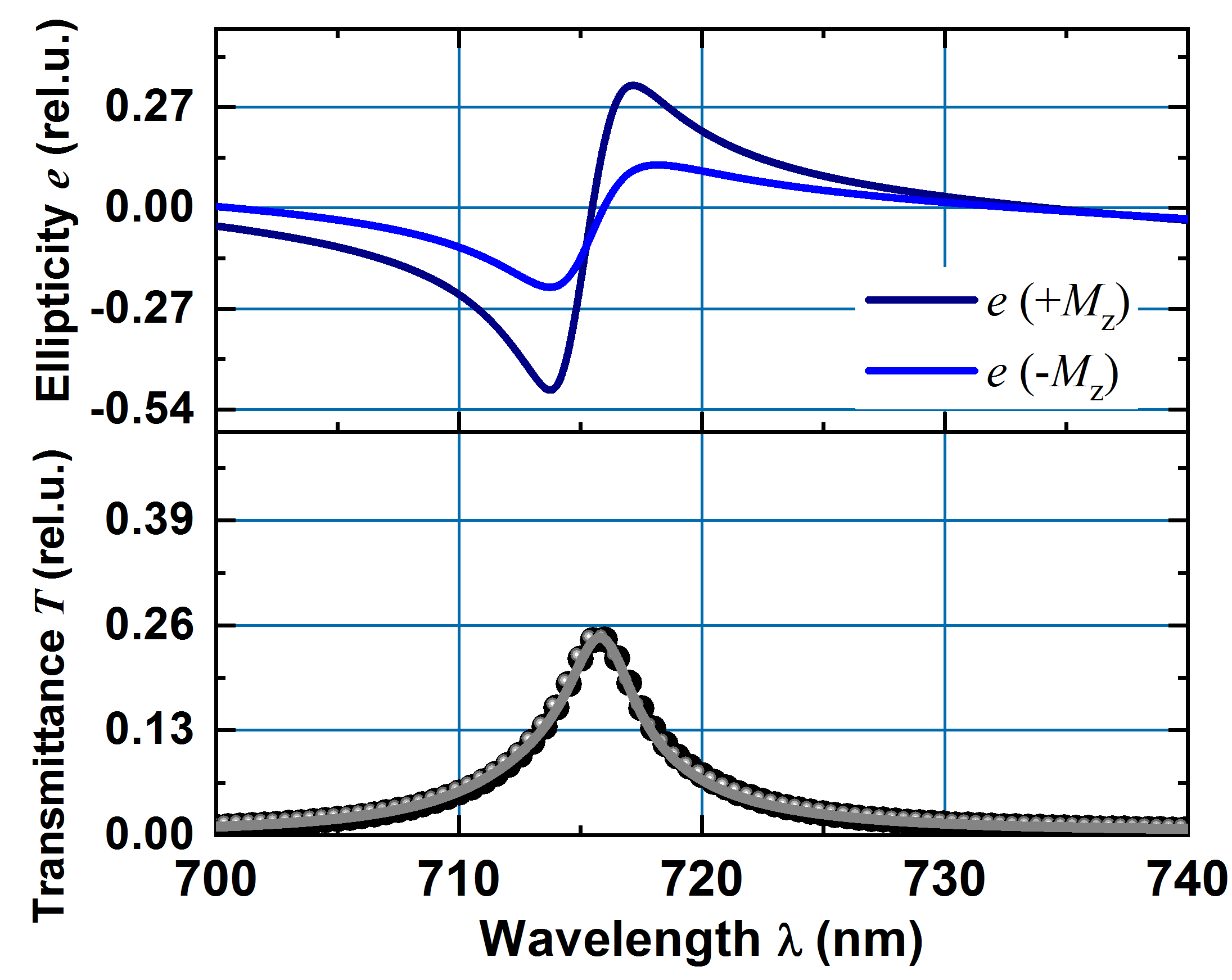}
\includegraphics[width=0.35\linewidth]{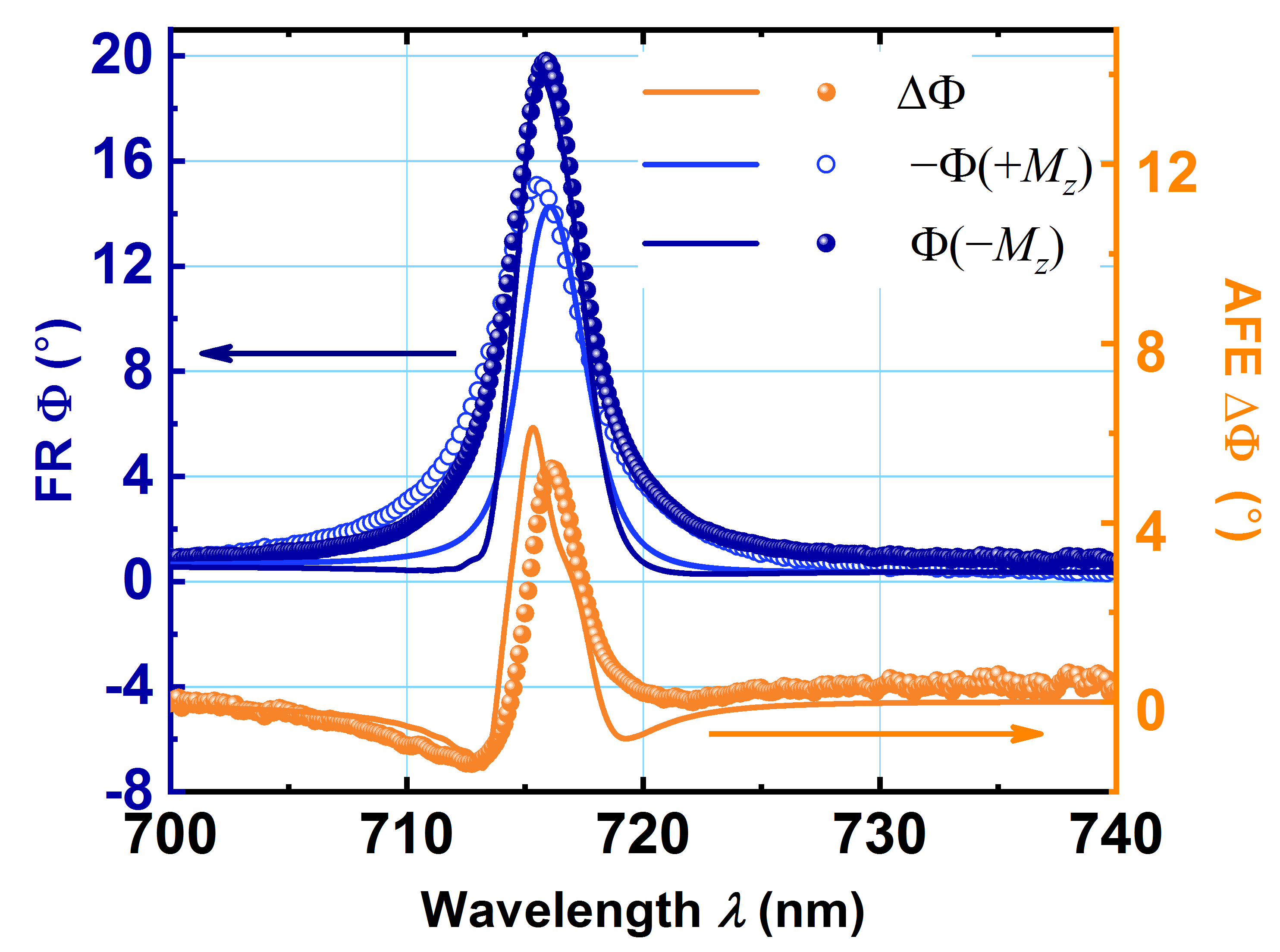}\\
\caption{(a) Schematic diagram of the investigated AFE in a MPC. (b) Optical ellipticity (top) and transmittance (bottom) spectra of the MPC for the oblique incidence $\theta=63^\circ$ of the light with $\Psi=70^\circ$ polarization angle. (c) The Faraday rotation angle spectra for the two opposite magnetizations and the asymmetric Faraday effect. Experimental data is presented by color spheres, while simulations are represented by solid curves.}
\label{fig: idea of AFE}
\end{figure*}

Such an impact of the spatial symmetry violation described by the director $\mathbf{L}$ might also be understood as the arisen non-equivalence between the clock- and counterclockwise rotation directions. In particular, such non-equivalence reveals itself as the difference between the two opposite directions of the magneto-optical rotations of linear polarization corresponding to the opposite magnetizations.

Let us illustrate how this asymmetry arising from the presence of a director $\mathbf{L}$ causes the non-equivalence between the clock- and counterclockwise directions of polarization rotation from the point of view of the structure's optical properties. Figure~\ref{fig: SI  clock rot} schematically shows the configuration characterized by a director $\mathbf{L}$ oriented along the $x$-axis (red arrow with the polar angle equal to $0^\circ$). The presence of a director causes the dependence of the optical characteristics of a structure $\Omega$ (which might be transmittance, absorption, etc.) on the polarization direction, as shown by the pseudo-color in Fig.~\ref{fig: SI  clock rot}. It is clearly seen that if the polarization of the incident light (white arrow) is parallel to $\mathbf{L}$ (Fig.~\ref{fig: SI  clock rot}a) then the rotations in clock- and counterclockwise directions are equivalent to each other since the opposite rotation angles bring the light to the states with the same $\Omega$ values (blue and green arrows, respectively). Due to this symmetry, $+\mathbf{M}$ and $-\mathbf{M}$ magnetizations induce the same magnitude of the Faraday polarization rotation. The same situation would be observed for $\mathbf{L}$ perpendicular to $\mathbf{E}$. In contrast to this, if the initial light polarization is tilted with respect to $\mathbf{L}$ (Fig.~\ref{fig: SI  clock rot}b), then the rotations in clock- and counterclockwise directions are not equivalent to each other. This might be seen in Fig.~\ref{fig: SI  clock rot}b, since the opposite rotation angles brings the light to the states with the different $\Omega$ values. Thus, the efficiency of the light interaction with a magnetic material is different in these cases. As a consequence, Faraday rotation would be different, as shown schematically by the blue and green arrows denoting polarizations $\mathbf{E}(\pm \mathbf{M})$. This is how the non-equivalence of counter- and clockwise Faraday rotations arises.

Non-equivalence of counter- and clockwise rotations, as well as the non-equivalence of the 'real-world' and 'mirror-world' configurations, are known as manifestations of the chirality phenomenon. The considerations made above show that such a chiral-type impact on the Faraday effect caused by the presence of the director $\mathbf{L}$ describing the structure symmetry could arise in non-chiral nanostructures and films.

It is important that such a symmetry break can be introduced by the design of a nanostructure or by incident light itself: an oblique incidence gives birth to the in-plane component of the light wavevector $\mathbf{k}_{\tau}$. In this case, $\mathbf{k}_{\tau}$ determines the orientation of $\mathbf{L}$ in the general consideration above. Such a configuration can be implemented in almost any type of magneto-optical structure, including smooth films and crystals. In the present work, we chose this option to introduce a symmetry break by the oblique incidence and experimentally demonstrate an asymmetric behavior of the Faraday rotation in a magnetophotonic crystal. After that, using the numerical simulations, we show that AFE arises in other magneto-optical configurations that justify the generality of the discussed phenomenon.

\section{Asymmetric Faraday effect in a magnetophotonic crystal}

\begin{figure*}[htb]
\centering
(a)~~~~~~~~~~~~~~~~~~~~~~~~~~~~~~~~~~~~~~~~~~~~~~~~~(b)~~~~~~~~~~~~~~~~~~~~~~~~~~~~~~~~~~~~~~~~~~~~~~~~~(c)\\
\includegraphics[width=0.6\columnwidth]{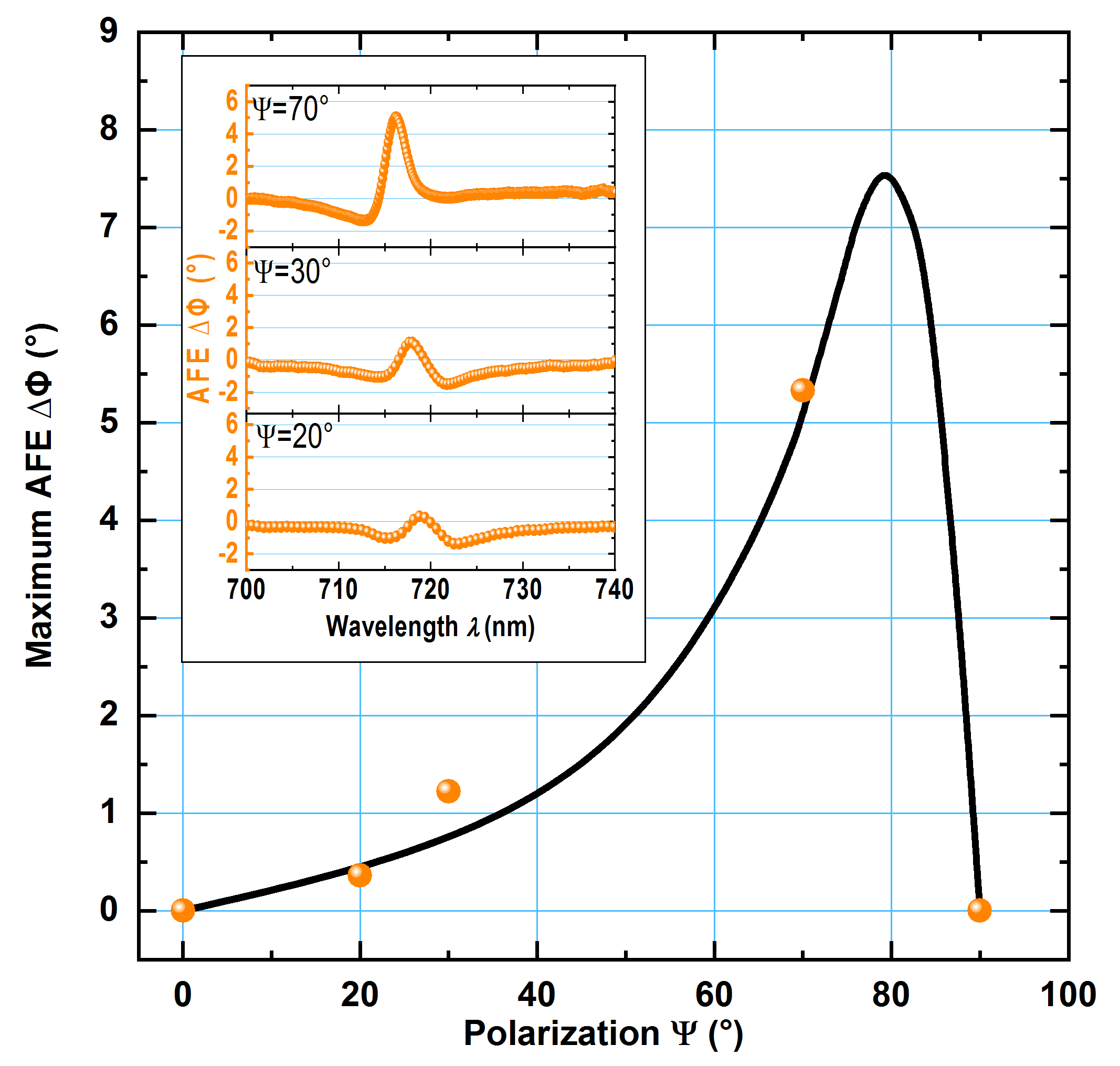}
\includegraphics[width=0.7\columnwidth]{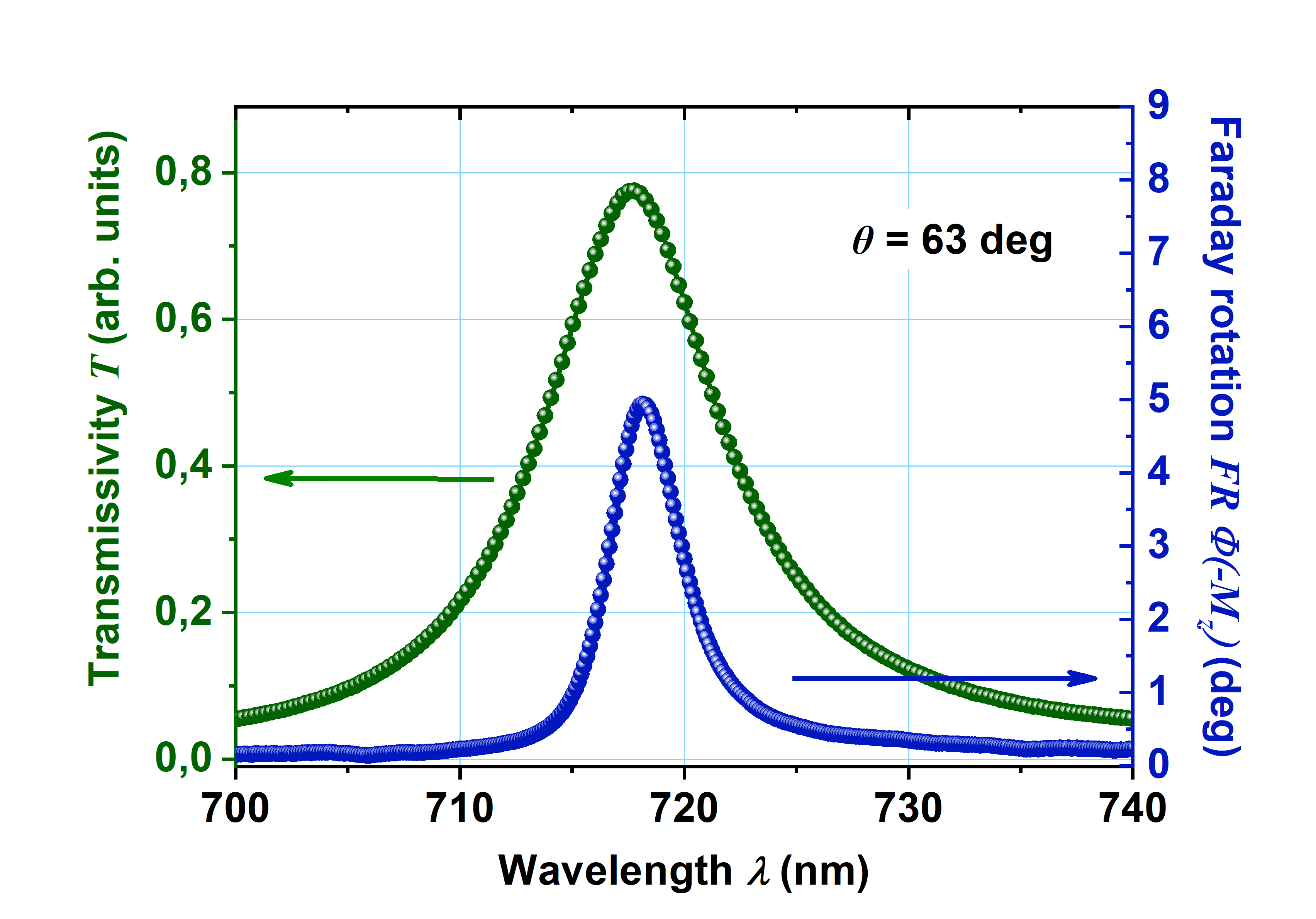}
\includegraphics[width=0.7\columnwidth]{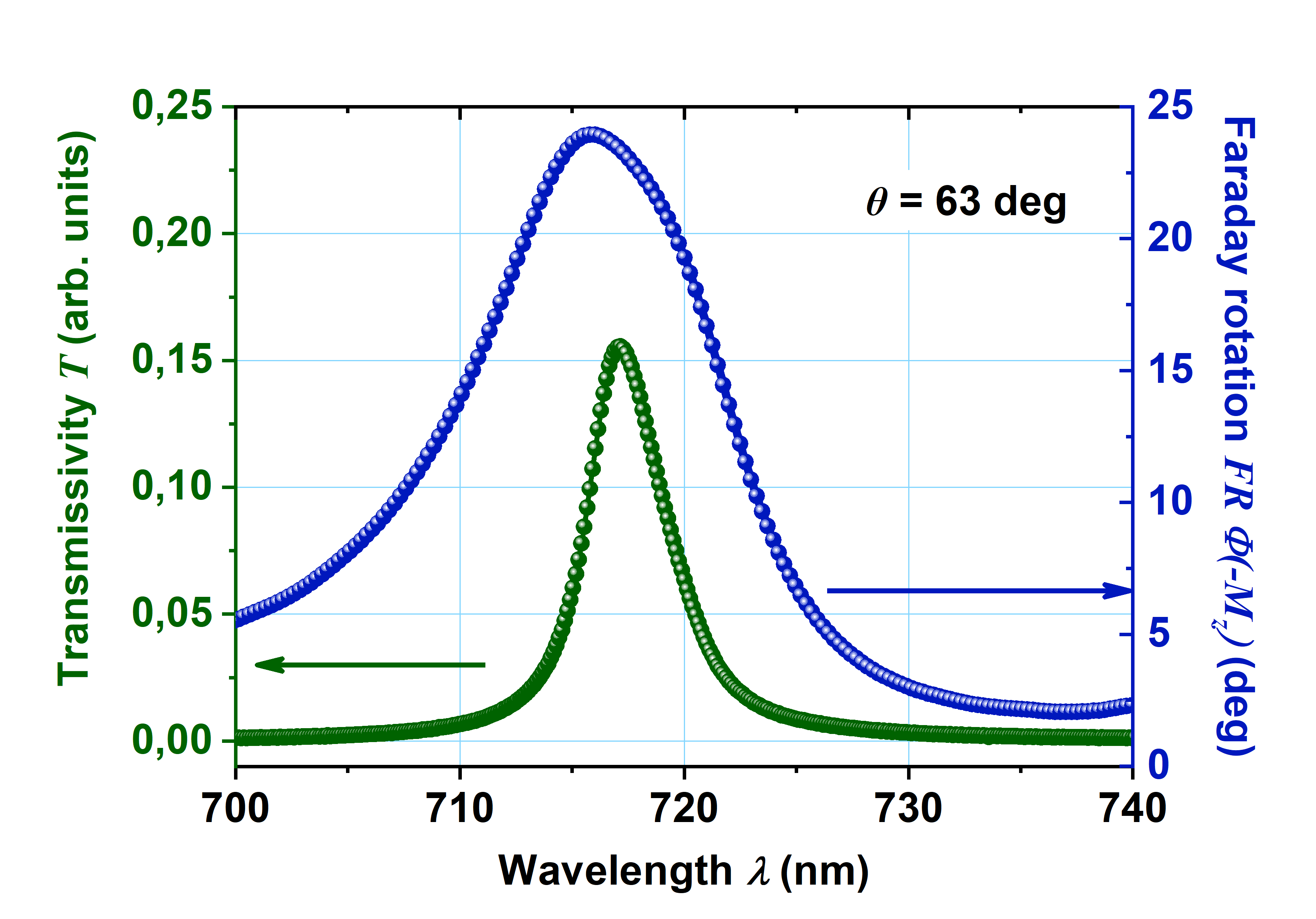}
\caption{(a) The dependence of the asymmetric Faraday effect on the initial light polarization $\Psi$ for $\theta$ = 63$^\circ$. The inset shows the AFE spectra for $\Psi=20^\circ, 30^\circ, 70^\circ$. Experimental data is presented by color spheres, while simulations – by solid curve. (b,c) Spectra of transmittance and Faraday rotation at the oblique incidence of $\theta = 63$ deg for (b) p- and (c) s-polarized light (experiments).}
\label{fig: T and Phi}
\end{figure*}

The magneto-photonic crystal (MPC) consists of two magnetic Bragg mirrors containing pairs of magnetic and nonmagnetic quarter-wave layers and a magnetic half-wave cavity layer sandwiched in between the mirrors (Fig.~\ref{fig: idea of AFE}a). It is illuminated by a linearly polarized light incident on the sample at the incidence angle $\theta$. Orientation of the light polarization is determined by the angle $\Psi$ between $\mathbf{E}$ and the plane of light incidence: $\Psi=0^\circ$ for the p-polarized light, and $\Psi=90^\circ$ for the s-polarized light (Fig.~\ref{fig: idea of AFE}a). Thus, $\Psi$ is the angle between the light polarization $\mathbf{E}$ and the director describing the magneto-optical configuration $\mathbf{L}$, which in this case coincides with $\mathbf{k}_{\tau}$. For the intermediate polarization states $0<\Psi<90^\circ$ a spatial symmetry break takes place and AFE is expected.

The asymmetric Faraday effect can be characterized by $\Delta \Phi$:
\begin{equation}
    \Delta \Phi = \Phi(+M_z)-(-\Phi(-M_z)) 
\end{equation}
where $\Phi(+M_z)$ and $\Phi(-M_z)$ are the Faraday rotation angles for opposite orientations of the magnetization vector $\mathbf{M}$ aligned parallel to the external magnetic field $\mathbf{H}$ along $z$ axis. Here we determine the Faraday angles as the difference between the total polarization rotation angles for the sample magnetized $\pm M_z$  and not magnetized $M_z=0$ along $z$-axis, therefore, purely optical polarization modification is omitted.

Experimental observation of the asymmetric Faraday effect was performed on the all-garnet MPC. The MPC Bragg mirrors consist of six pairs of diamagnetic $\mathrm{Sm_3Ga_5O_{12}}$ (SGG) and the MO active ferrimagnetic $\mathrm{Bi_{2.97}Er_{0.03}Al_{0.5}Ga_{0.5}O_{12}}$ (BIG) garnet layers ($h_\mathrm{BIG} = 75$~nm, $h_\mathrm{SGG} = 100$~nm) synthesized by RF-magnetron sputtering on (111) $\mathrm{Gd_3Ga_5O_{12}}$ (GGG) substrate~\cite{ansari2012multicolor,passler2017generalized}. There is a cavity layer of BIG in between the Bragg mirrors: GGG / [BIG / SGG]$^6$ /2 BIG / [SGG / BIG]$^6$ responsible for a cavity mode (see Supplementary S2 and S3 for the details).

The Faraday rotation was experimentally found using a conventional scheme where the transmitted light intensity $I_\mathrm{out}$ is measured in the presence of a polarizer and analyzer crossed by an angle of $45^\circ$, so that  $I_\mathrm{out}=\frac{1}{2}I_0 \cdot (1-\sin(2\mathrm{OR} + 2\Phi)$, where $\mathrm{OR}$ is a purely optical non-magnetic polarization rotation, and $I_0$ is the intensity of the light transmitted through the MPC without a polarizer (see Supplementary S3.3 for the details). Strictly speaking, this formula is not valid if the light ellipticity arises. But still, it can be applied  in the vicinity of the cavity resonance since the ellipticity vanishes at the 
resonant wavelength (Fig.~\ref{fig: idea of AFE}b).

We chose at first the polarization angle $\Psi=70^\circ$. Both experimental measurements (color spheres in Fig.~\ref{fig: idea of AFE}b) and numerical simulations performed using the transfer matrix method 4×4~\cite{li2003fourier} (solid curves in Fig.~\ref{fig: idea of AFE}b, see Supplementary S2 for simulation details) show that at the oblique incidence, the cavity mode is observed at $\lambda=716$~nm for $\theta = 63^\circ$. The cavity resonance is accompanied with a pronounced peak of the Faraday rotation $\Phi\sim 20^\circ$ (Fig.~\ref{fig: idea of AFE}c), which gives a high magneto-optical quality factor $Q_\mathrm{MO}= -2 |\Phi| / \ln T = 42.6^\circ$ and MO figure of merit $\mathrm{MOFOM} = T \sin2\Phi = 20\%$~\cite{mikhailova2018optimization}.

At $\Psi=70^\circ$ a notable AFE characterized by $\Delta \Phi=5.5^\circ$ appears in the vicinity of the cavity resonance (Fig.~\ref{fig: idea of AFE}c). Since at this wavelength $\Phi\sim20^\circ$, it means that AFE accounts about 30$\%$ of the Faraday effect. Similar AFE is observed for oblique incidence of any intermediate polarization of light and vanishes only for pure p- or s-polarizations for which there is no spatial symmetry violation since in those cases $\mathbf{k}_{\tau}$ is either parallel or perpendicular to $\mathbf{E}$, respectively (Fig.~\ref{fig: T and Phi}a). The largest value of AFE appears at $\Psi=80^\circ$ and equals to $\Delta \Phi=7.5^\circ$, which is $38\%$ of the $\Phi$ value.

From a theoretical point of view, the asymmetry of the Faraday effect corresponds to the emergence of even in magnetization terms of the Faraday rotation in the expansion: $\Phi(M_z)=\phi_1\frac{M_z}{M_{z0}}+\phi_2\left(\frac{M_z}{M_{z0}}\right)^2+...$ (where $M_{z0}$ is the saturation magnetization and $\phi_j$ are the expansion coefficients). In the 'usual' cases of normal incidence, or pure p- or s-polarized light at the oblique incidence, which are used in most of the magneto-optical experiments, $\phi_{2}=\phi_{2n}=0$, where $n$ is an integer (see Supplementary S1.2). 

\begin{figure*}[htb]
\centering
(a)~~~~~~~~~~~~~~~~~~~~~~~~~~~~~~~~~~~~~~~~~~~~~~~~~~~~~~~~~~~~~~~~~~~~~~~~~~(b)\\
\includegraphics[width=0.8\linewidth]{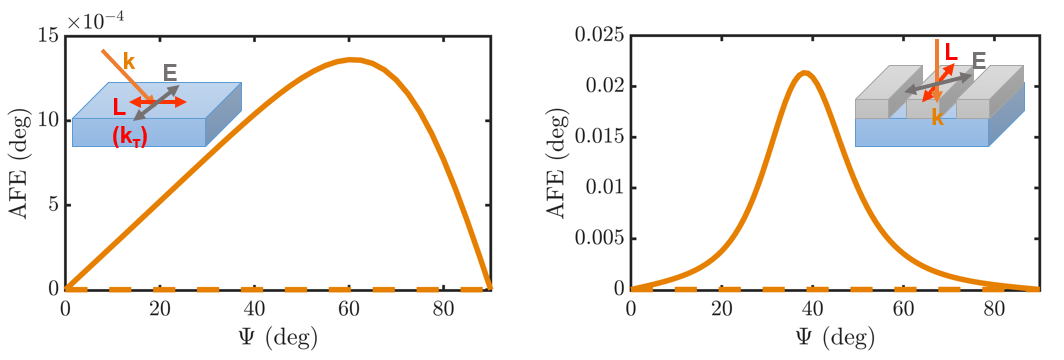}
\caption{Asymmetric Faraday effect caused by the spatial symmetry break. (a) Symmetry break by light: AFE arising for the oblique (solid line) and normal (dashed line) incidence on a smooth iron-garnet film on a $\mathrm{SiO_2}$ substrate. (b) Symmetry break caused by the structure: AFE arising at normal incidence in a smooth iron-garnet film covered by 1D Si grating (solid line) or smooth Si layer (dashed line).}
\label{fig: AFE var struct}
\end{figure*}

As AFE is a nonlinear part of the Faraday rotation, one might expect to observe high AFE in cases where the linear Faraday effect is large. From this point of view, MPCs at oblique incidence are the ideal candidates to observe AFE. On the one hand, they possess a high Faraday rotation of tens of degrees (see blue lines in Figs.~\ref{fig: T and Phi}b,c). On the other hand, they exhibit a prominent polarization dependence of the optical properties at the incline incidence. The origin of this dependence is the variation of the reflectivity of the Bragg mirrors surrounding the cavity layer, leading to a difference in Q-factors for different light polarizations. Figure~\ref{fig: T and Phi} demonstrates a more than 5-fold difference in the transmittance (green lines) and Faraday rotation (blue lines) for p-polarized (Fig.~\ref{fig: T and Phi}b) and s-polarized (Figure~\ref{fig: T and Phi}c) light (see also Supplementary S3.2). Thus, MPC properties are very sensitive to the mutual orientations of $\mathbf{E}$ and $\mathbf{k}_\tau$ vectors, which clearly differentiates $\Psi+\Phi$ and $\Psi-\Phi$ states corresponding to the opposite rotation directions (see also a sketch in Fig.~\ref{fig: SI  clock rot}). This provides large values of AFE. For the considered MPC, the second-order term contribution grows significantly up to $\phi_2/\phi_1=0.19$. For a comparison, in a smooth film of the same thickness, the quadratic term is very small: $\phi_2/\phi_1=0.009$ (see below). Therefore, the role of the MPC here is to enhance the AFE and make it easily measurable. 

The Faraday effect is commonly used for the magneto-optical studies of the magnetic materials and magnetization distributions. However, the obvious difficulties arise if the magnetization distribution is non-uniform so that the net magnetization in the area illuminated by the probe beam is zero: $\int_{S_{\mathrm{beam}}} M_{z}(x,y) \mathrm{d}x \mathrm{d}y= 0$. This situation takes place, for example, for macroscopic static subwavelength domain structures~\cite{danneau2002individual} and dynamic $ M_{z}$ oscillations in ultrashort spin waves~\cite{dieterle2019coherent,sluka2019emission} and spin-wave resonances~\cite{chernov2020all}. Although there is magnetization in each spatial point of the material, the Faraday rotation is absent due to the zero average $\left<M_z\right>$ in the illuminated area. The situation changes if the spatial symmetry break is used to produce AFE, for example, as demonstrated in the present work. Even if the net magnetization is zero, it is possible to measure the polarization rotation arisen due to the AFE. This opens a unique possibility to study various types of materials with inhomogeneous static or dynamic spatial magnetization distributions. 

\section{Asymmetric Faraday effect in other structures}

The symmetry break described by a director characterizing the magneto-optical configuration is a quite general condition. This means that AFE arises in a wide range of configurations with the discussed symmetry break produced either by the light itself, or by the structure. This statement is justified by the numerical simulations of the two most simple configurations. 

Figure~\ref{fig: AFE var struct}a demonstrates the AFE arising due to the symmetry break introduced by the light oblique incidence with $\mathbf{k}_\tau$ vector on a smooth iron-garnet film. This situation is similar to the MPC discussed above, but the configuration is more simple. AFE arises for the intermediate polarizations where $\mathbf{E}$ and $\mathbf{k}_\tau$ are tilted with respect to each other (Fig.~\ref{fig: AFE var struct}a, solid line). As it was theoretically predicted, AFE disappears for pure p- and s-polarized light ($\Psi=0^\circ$ and $\Psi=90^\circ$ points of a solid line in Fig.~\ref{fig: AFE var struct}a) where the vectors $\mathbf{k}_\tau$ and $\mathbf{E}$ are either orthogonal or collinear to each other. Also, AFE vanishes at the normal incidence when $\mathbf{k}_\tau=0$ (dashed line in Fig.~\ref{fig: AFE var struct}a). 

On the contrary, Fig.~\ref{fig: AFE var struct}b demonstrates the AFE arising due to the symmetry break introduced by the structure. 1D Si grating is placed on top of the smooth iron-garnet film. This configuration can be described by a director $\mathbf{L}$ oriented along the stripes. Thus, even for a normal incidence, if polarization is tilted with respect to $\mathbf{L}$, i.e. in the case of polarizations intermediate between the parallel and perpendicular to the Si grating stripes, AFE is observed (solid line in Fig.~\ref{fig: AFE var struct}b). AFE vanishes in the cases of $\mathbf{E}$ parallel and perpendicular to $\mathbf{L}$ ($\Psi=0^\circ$ and $\Psi=90^\circ$ points of a solid line in Fig.~\ref{fig: AFE var struct}b). AFE is also absent if Si grating is transformed into the continuous Si layer and therefore a preferred direction described by $\mathbf{L}$ disappears (dashed line in Fig.~\ref{fig: AFE var struct}b).

\section{Conclusion}

Therefore, here we demonstrate the asymmetric Faraday effect that arises if an additional symmetry break with respect to the reflection in a mirror parallel to magnetization is introduced to the magneto-optical system. We confirm this prediction by studying the magneto-photonic crystal at oblique incidence, which provides a unique modification of the Faraday effect's symmetry and magnetization dependence. Experimentally, a pronounced asymmetry of the Faraday rotation spectra for oppositely directed magnetic fields reaching $|\Phi(+M_z)|-|\Phi(-M_z)|\sim30\%$ is obtained. Such behavior results from a strong non-equivalence of the counter- and clockwise magneto-optical rotations of linear polarization and brings the quadratic in the magnetization term to the Faraday rotation spectra. Breaking the oddness of the Faraday effect provides a unique possibility to observe magneto-optical polarization rotation even in materials with non-uniform magnetization direction and zero net magnetization. At the same time, we show that AFE is a quite general effect that might be observed in various structures with a particular type of spatial symmetry breaking. This is a key feature for optical studies of static and dynamic magnetic patterns with zero net magnetization and ultrashort spin waves~\cite{dieterle2019coherent,sluka2019emission}.

\begin{acknowledgments}
This work was financially supported by Russian Science Foundation, project No. 24-42-02008.
The authors thank A.N. Kalish and A.A.Voronov for the fruitful discussion.
\end{acknowledgments}

\bibliography{apssamp}

\end{document}



\title{Supplementary Information\\
Asymmetric Faraday effect caused by a break of spatial symmetry}

\author{D.O. Ignatyeva}
\affiliation{Institute of Physics and Technology, V.I. Vernadsky Crimean Federal University, 295007 Simferopol, Crimea}
\affiliation{Photonic and Quantum technologies school, Faculty of Physics, Lomonosov Moscow State University, Leninskie gori, 119991 Moscow, Russia}
\affiliation{Russian Quantum Center, 121205 Moscow, Russia}

\author{T.V. Mikhailova}
\affiliation{Institute of Physics and Technology, V.I. Vernadsky Crimean Federal University, 295007 Simferopol, Crimea}

\author{P.O. Kapralov}
\affiliation{Russian Quantum Center, 121205 Moscow, Russia}

\author{S.D. Lyashko}
\affiliation{Institute of Physics and Technology, V.I. Vernadsky Crimean Federal University, 295007 Simferopol, Crimea}

\author{V.N. Berzhansky}
\affiliation{Institute of Physics and Technology, V.I. Vernadsky Crimean Federal University, 295007 Simferopol, Crimea}

\author{V.I. Belotelov}
\affiliation{Institute of Physics and Technology, V.I. Vernadsky Crimean Federal University, 295007 Simferopol, Crimea}
\affiliation{Photonic and Quantum technologies school, Faculty of Physics, Lomonosov Moscow State University, Leninskie gori, 119991 Moscow, Russia}
\affiliation{Russian Quantum Center, 121205 Moscow, Russia}

\date{\today}

\maketitle

\tableofcontents

\section{Magneto-optical tensor and effects}

\subsection{Permittivity tensor and normal-mode equation for the magneto-optical media}

Magneto-optical properties of the isotropic magnetic material could be described by the following constitutive equations~\cite{zvezdin1997modern}:
\begin{equation}
    \mathbf{D}=\varepsilon_0\mathbf{E} + i \left[\mathbf{g}\times \mathbf{E} \right] + b \left(\mathbf{E} - \mathbf{m}(\mathbf{m} \cdot \mathbf{E})\right),
\end{equation}
where $\varepsilon_0$ is a permittivity of a non-magnetized medium, $\mathbf{m}=\mathbf{M}/M$ is a unitary vector co-directed with medium magnetization, $M=|\mathbf{M}|$ is magnetization absolute value, $\mathbf{g}=\alpha \mathbf{M}$ is a gyration vector, $b(M)=\beta \mathbf{M}^2 = \varepsilon_M-\varepsilon_0$ is the quadratic in magnetization coefficient. This corresponds to the Hermitian type of the permittivity tensor $\varepsilon_{ij}=\varepsilon_{ji}^*$, which for the magnetization along the z-axis has the form:
\begin{equation}
\hat{\varepsilon}=
\begin{pmatrix}
\varepsilon_M & ig & 0 \\
-ig & \varepsilon_M & 0 \\
0 & 0 & \varepsilon_0 \\
\end{pmatrix}.
\end{equation}
According to the Onsager principle, antisymmetric $\mathbf{g}$ and symmetric $b(M)$ magneto-optical contributions to the permittivity tensor are odd and even with respect to time reversal, and, consequently, linear and quadratic in medium magnetization~\cite{kalashnikova2015ultrafast}.

Depending on the mutual orientation of light $\mathbf{E}$, $\mathbf{k}$ vectors and medium magnetization $\mathbf{M}$, both odd and even in magnetization effects could be observed. These effects are described by the normal-mode equation, which is directly derived from Maxwell's equations:
\begin{equation}
    n^2\mathbf{E}-\mathbf{n}(\mathbf{n} \cdot \mathbf{E})= \mathbf{D},
    \label{Eq nmode}
\end{equation}
where $\mathbf{n}=\mathbf{k}/k_0$ is the refractive index vector. 

\subsection{The Faraday rotation: odd magneto-optical effect}
The Faraday magneto-optical effect arises when the light travels along the magnetization direction, $\mathbf{m}\parallel \mathbf{k}$. The normal mode equation~\eqref{Eq nmode} has a solution in the form of the two circularly polarized modes, $\sigma^+$ and $\sigma^-$, with the different refractive indices:
\begin{equation}
    n^2_{\pm}=\varepsilon_M \left(1 \pm \frac{g}{\varepsilon_M} \right).
\end{equation}
The difference between the refractive indices of $\sigma^+$ and $\sigma^-$ causes the rotation of the linearly polarized light propagating a distance $L$ in such a medium:
\begin{align}
   \Phi=-\frac{1}{2}k_0 L (n_+-n_-)=~~~~~~~~~~~~~~~~~~~~~~~~~~~~\nonumber\\
   =-\frac{1}{2}k_0 L \sqrt{\varepsilon_M}\left(\sqrt{1  + \frac{g}{\varepsilon_M}}-\sqrt{1 -\frac{g}{\varepsilon_M} }\right).
\end{align}

The gyration coefficient $g$ is usually much smaller than permittivity; for example, it is about $g\sim 0.01 ... 0.001$ for iron-garnets of different compositions~\cite{zvezdin1997modern} while the typical permittivity is in the range $\varepsilon_0=4.5...6.5$. For the considered iron-garnet at $\lambda=715$~nm wavelength $\varepsilon_0=6.858+0.014i$, and $g=0.021$ (see Sec.~\ref{Sec params}). Thus, one can make the Taylor-series expansion:
\begin{align}
\Phi=~~~~~~~~~~~~~~~~~~~~~~~~~~~~~~~~~~~~~~~~~~~~~~~~~~~~~~~~~~~~~~~~~\nonumber\\
    -\frac{1}{2}k_0 L\biggl( \sqrt{\varepsilon_0}+\frac{b}{2\sqrt{\varepsilon_0}}\biggr)\biggl(1 + \frac{1}{2}\frac{g}{\varepsilon_M}- \frac{1}{8}\frac{g^2}{\varepsilon_M^2} +\frac{1}{16}\frac{g^3}{\varepsilon_M^3}-...\nonumber\\
    - 1 + \frac{1}{2}\frac{g}{\varepsilon_M} + \frac{1}{8}\frac{g^2}{\varepsilon_M^2} +\frac{1}{16}\frac{g^3}{\varepsilon_M^3}+...\biggr) = \nonumber \\
    =-\frac{1}{2}k_0 L \biggl( \sqrt{\varepsilon_0}+\frac{b}{2\sqrt{\varepsilon_0}}\biggr) \biggl( \frac{g}{\varepsilon_M} +\frac{1}{8}\frac{g^3}{\varepsilon_M^3} + ...\biggr)= \nonumber\\
    =-\frac{1}{2}k_0 L \biggl( \sqrt{\varepsilon_0}+\frac{b}{2\sqrt{\varepsilon_0}}\biggr) \biggl( \frac{g}{\varepsilon_0} - \frac{g b}{\varepsilon_0^2} +\frac{1}{8}\frac{g^3}{\varepsilon_0^3} + ...\biggr) = \nonumber\\
    =-\frac{1}{2}k_0 L \frac{g}{\sqrt{\varepsilon_0}}\biggl( 1 -\frac{b}{2\varepsilon_0} + \frac{1}{8}\frac{g^2}{\varepsilon_0^2} + ...\biggr) =\nonumber\\ 
    =-\frac{1}{2}k_0 L \frac{\alpha}{\sqrt{\varepsilon_0}}\biggl( M -\frac{\beta M^3}{2\varepsilon_0} + \frac{1}{8}\frac{\alpha^2 M^3}{\varepsilon_0^2} \biggr) .\label{Phi long}
    \end{align}
The obtained formula illustrates the well-known property of the Faraday effect, which is odd in magnetization direction. Even when higher orders of $M$ expansion terms are considered, it is clear that $\Phi$ does not contain any even in magnetization terms. The next non-zero term in $\Phi$ is proportional to $M^3$ and is $(-\beta/2\varepsilon_0 + \alpha^2/8\varepsilon^2)\sim 10^{-6}$ times smaller than the first linear one and definitely can be neglected. The oddness of the Faraday effect in the magnetization is a consequence of the time-reversal symmetry. The time reversal simultaneously changes the magnetization direction $\mathbf{m}\rightarrow-\mathbf{m}$ and the light helicity $\sigma^+\rightarrow\sigma^-$.

\subsection{Magnetic linear birefringence: even magneto-optical effect} \label{Sec in-plane}
The magnetic linear birefringence arises if the medium is magnetized perpendicular to the light propagation direction: $\mathbf{m}\perp\mathbf{k}$.
In this case, normal mode equation~\eqref{Eq nmode} has the solution in the form of the two linearly polarized waves, $\mathbf{E}\parallel \mathbf{m}$ and $\mathbf{E}\perp\mathbf{m}$ with different refractive indices:
\begin{align}
    n_{\parallel}^2&=\varepsilon_0,\label{n_perp}\\
    n_{\perp}^2&=\varepsilon_M -\frac{g^2}{\varepsilon_0}.
    \label{n_par}
\end{align}
The difference between the refractive indices of these two linearly polarized waves is:
\begin{equation}
    n_{\perp}-n_{\parallel}= \frac{1}{2\sqrt{\varepsilon_0}}\biggl(b-\frac{g^2}{\varepsilon_0}\biggr)=\frac{1}{2\sqrt{\varepsilon_0}}\biggl(\beta-\frac{\alpha^2}{\varepsilon_0}\biggr)M^2
\end{equation}
Notice that $n_{\perp}-n_{\parallel}$ contains only even in magnetization terms. This can be understood on the basis of time reversal which results in $\mathbf{m}\rightarrow\mathbf{-m}$, but both of these directions are equivalent from the point of view of linearly polarized light with $\mathbf{k}\perp\mathbf{m}$. Namely, this configuration corresponds to the orientational magneto-optical effect which arises from the Fresnel coefficients difference for $\mathbf{E}\parallel \mathbf{m}$ and $\mathbf{E}\perp\mathbf{m}$ polarizations, and Coutton-Mouton effect which reveals itself in appearance of the magneto-optical ellipticity similar to the case of the natural birefringence. It can be clearly seen that both effects related to such configuration are $\sim10^{-3}$ times smaller than the ones observed for a Faraday configuration.

An important consequence of Eq.~\eqref{n_perp},~\eqref{n_par} is that the magnetization of the sample perpendicular to the light plane of incidence (which is also called the transverse configuration) does not produce the magneto-optical polarization rotation of light. Thus, this configuration is used to determine the purely optical contribution to the rotation of the mixed polarization state which arises due to the Fresnel reflection at the interfaces of the magnetophotonic crystal layers. 

\section{Material parameters and simulations}
\subsection{Numerical simulations via transfer matrix method}

Numerical simulations of the optical and magneto-optical spectra of the magnetophotonic crystal were performed by numerical solution of Maxwell's equations by
the transfer matrix method $4\times4$. The details on this method are provided in~\cite{vivsvnovsky1995polar}. Each of the layers of the magnetophotonic crystal was described by the permittivity tensor, so that $\varepsilon_{ij}=i\epsilon_{ijk}g_k$~\cite{zvezdin1997modern}, where $\epsilon_{ijk}$ is a Levi-Civita symbol and $g_k$ are the components of the gyration vector that is parallel to the magnetization of the material. Actually, $g_k$ was non-zero only for ferrimagnetic $\mathrm{Bi_{2.97}Er_{0.03}Al_{0.5}Ga_{0.5}O_{12}}$ (BIG) garnet. 

The wavelength dependence of all of the permittivity tensor components was taken into account, as well as the Fresnel reflection from the backside of a transparent substrate. The exact values of these components provided in Sec.~\ref{Sec params}.

The transfer matrix method provides the optical characteristics of the structure, such as transmittance, reflectance, absorption, polarization rotation, and others. 

\subsection{Material parameters} \label{Sec params}

\begin{figure*}[htb]
\centering
\includegraphics[width=0.45\linewidth]{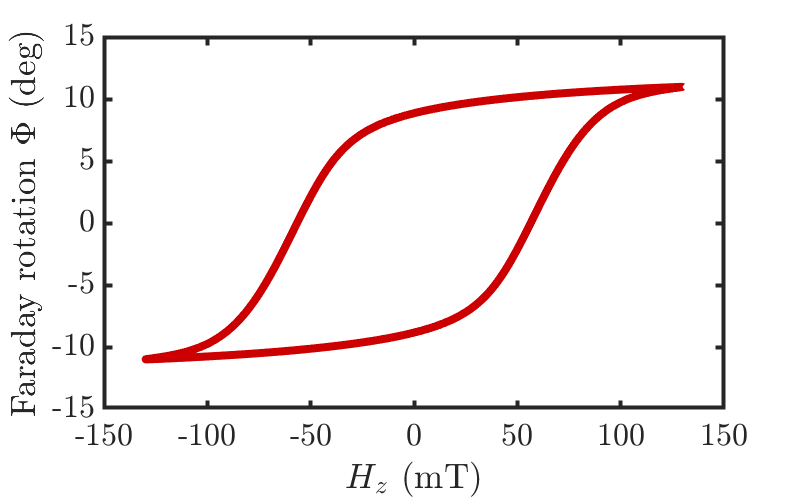}
\includegraphics[width=0.45\linewidth]{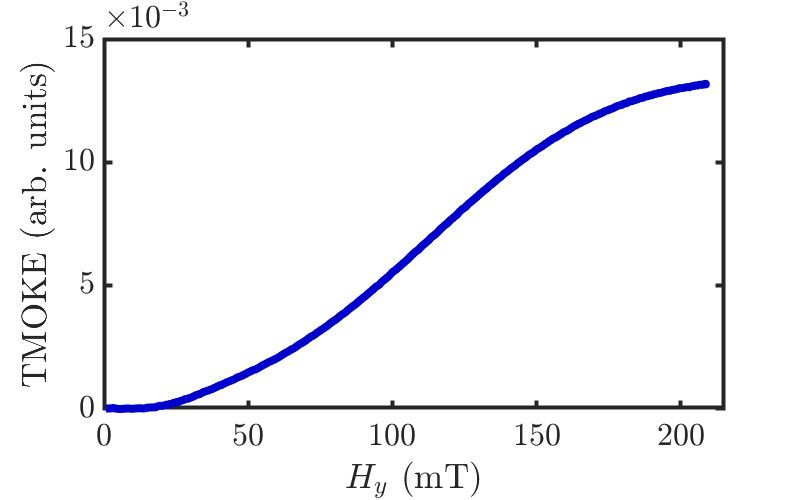}
\caption{Experimental hysteresis loops obtained for the magnetophotonic crystal.}
\label{fig: Hyst Far TMOKE}
\end{figure*}

The following wavelength-dependent material parameters were used for the simulations. The refractive index of (111) $\mathrm{Gd_3Ga_5O_{12}}$ (GGG) substrate is equal to
\begin{equation}
    n_\mathrm{GGG}=1.91+\left(\frac{203 \left[\mathrm{nm}\right]}{\lambda}\right)^2,
\end{equation}
and the refractive index of air is $n_\mathrm{air}=1$. 

The refractive index of diamagnetic $\mathrm{Sm_3Ga_5O_{12}}$ (SGG) material $n_\mathrm{SGG}=\sqrt{\varepsilon_\mathrm{SGG}}$ has the form:
\begin{equation}
    \varepsilon_\mathrm{SGG}=1+\frac{2.75}{1-\left(\frac{128 [\mathrm{nm}]}{\lambda}\right)^2};
\end{equation}

The refractive index of the magneto-optical ferrimagnetic $\mathrm{Bi_{2.97}Er_{0.03}Al_{0.5}Ga_{0.5}O_{12}}$ (BIG) garnet is:
\begin{align}
    n_\mathrm{BIG}=1+\frac{4.70}{1-
    \left(\frac{303 [\mathrm{nm}]}{\lambda}\right)^2} &+ \nonumber\\
    +\frac{0.08}{1-\left(\frac{494 [\mathrm{nm}]}{\lambda}\right)^2}+&
    i\cdot 0.07 \cdot \frac{494 [\mathrm{nm}]}{\lambda},
\end{align}
while its magneto-optical activity is described by the gyration coefficient:
\begin{align}
    g=0.40 - 1.85 \cdot 10^{-3} [\mathrm{nm}^{-1}]\lambda + 3.23\cdot10^{-6} [\mathrm{nm}^{-2}] \lambda^2 - \nonumber \\
    - 2.49\cdot10^{-9} [\mathrm{nm}^{-3}] \lambda^3 + 7.22\cdot10^{-13} [\mathrm{nm}^{-4}]\lambda^4. 
\end{align}

These values were determined from the optical and magneto-optical spectra of the samples and agree with the ones reported for the garnets with similar compositions~\cite{grishin2012luminescent,grishin2019waveguiding}.

\section{Experimental measurements}
\subsection{Magneto-optical hysteresis loops}

All measurements were performed at room temperature. First, the magneto-optical hysteresis loops of the Faraday effect at normal incidence were measured (Fig.~\ref{fig: Hyst Far TMOKE}a). The out-of-plane coercive magnetic field needed to saturate the sample is $H_{zc}=120~$mT. This agrees well with the previous results obtained for similar materials~\cite{grishin2012luminescent}. 

Sample characterization for the in-plane configuration of the applied external magnetic fields was performed using the transverse magneto-optical Kerr effect (TMOKE) measurements for the oblique incidence of p-polarized light (Fig.~\ref{fig: Hyst Far TMOKE}b). The TMOKE value was measured as the reflected light intensity change for the magnetic field changing from $+H_y$ to $-H_y$, so that $\mathrm{TMOKE}=I(+H_y)-I(-H_y)$. The in-plane coercive magnetic field needed to saturate the sample is $H_{yc}=200~$mT.

In our further experiments, we used the magnetic fields exceeding these values, $H_z$=220~mT and $H_y$=370~mT, correspondingly, in order to ensure that magnetization was aligned with the external magnetic field.

\subsection{MPC optical and magneto-optical spectra}
\begin{figure*}[htb]
\centering
(a)~~~~~~~~~~~~~~~~~~~~~~~~~~~~~~~~~~~~~~~~~~~~~~~~~~~~~~~~~(b)\\
\includegraphics[width=0.4\linewidth]{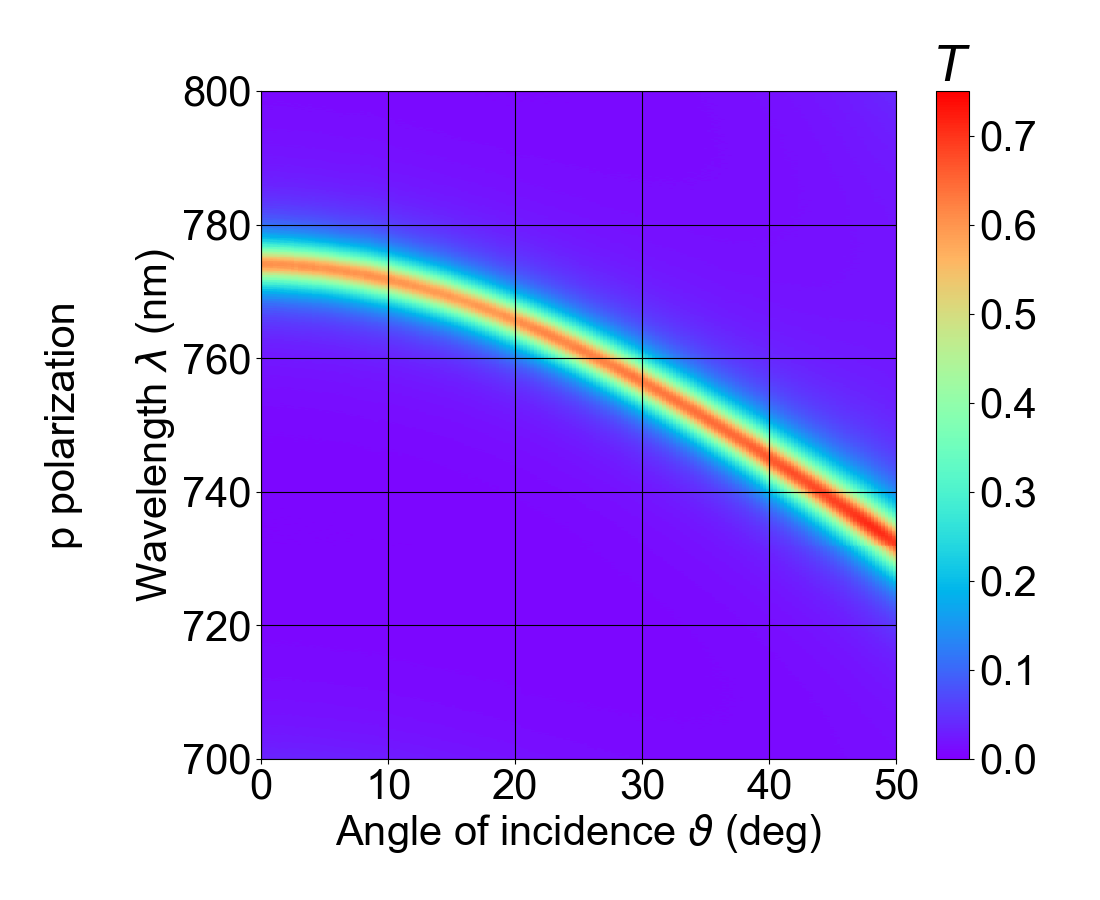}
\includegraphics[width=0.4\linewidth]{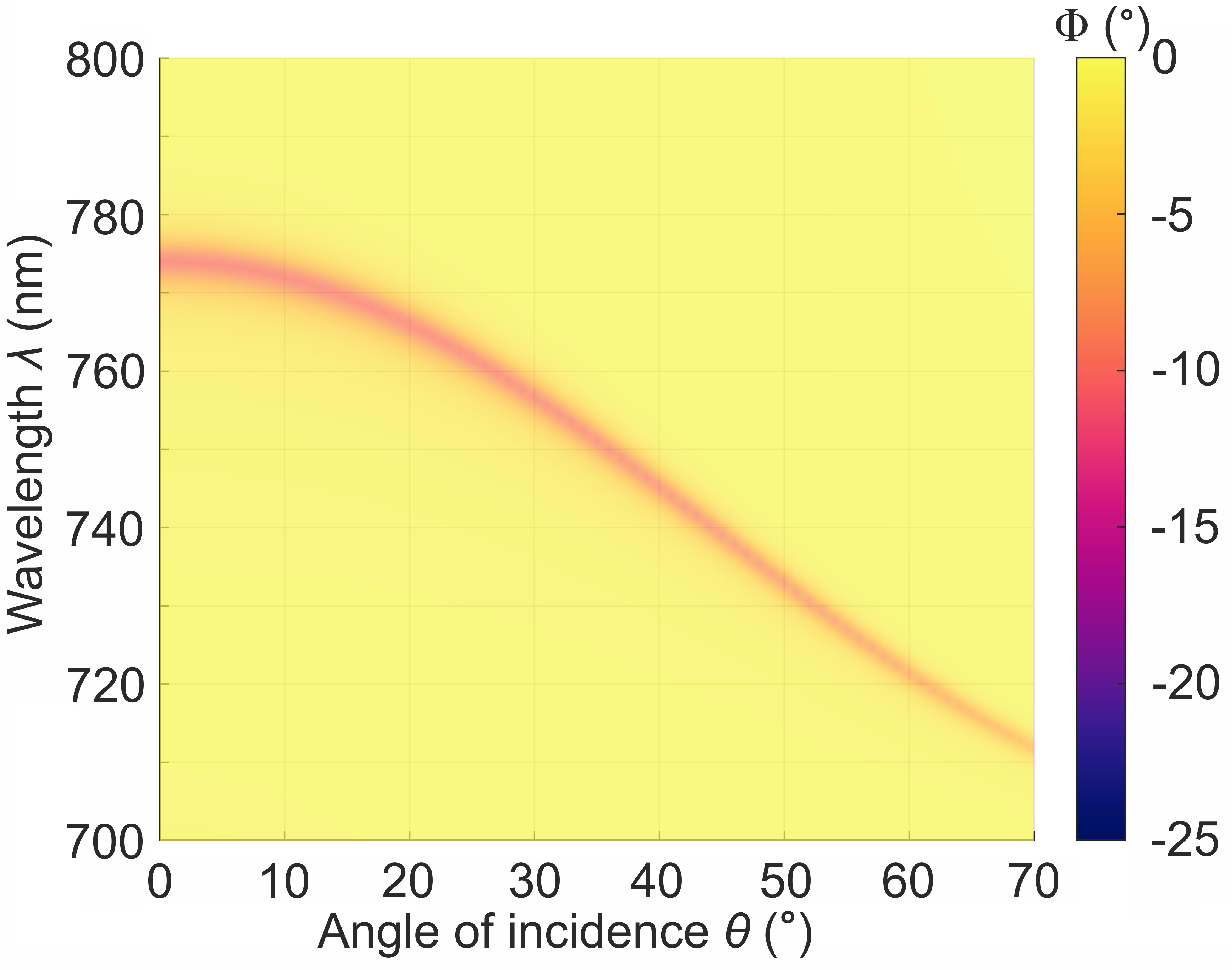}\\~\\
(c)~~~~~~~~~~~~~~~~~~~~~~~~~~~~~~~~~~~~~~~~~~~~~~~~~~~~~~~(d)\\
\includegraphics[width=0.4\linewidth]{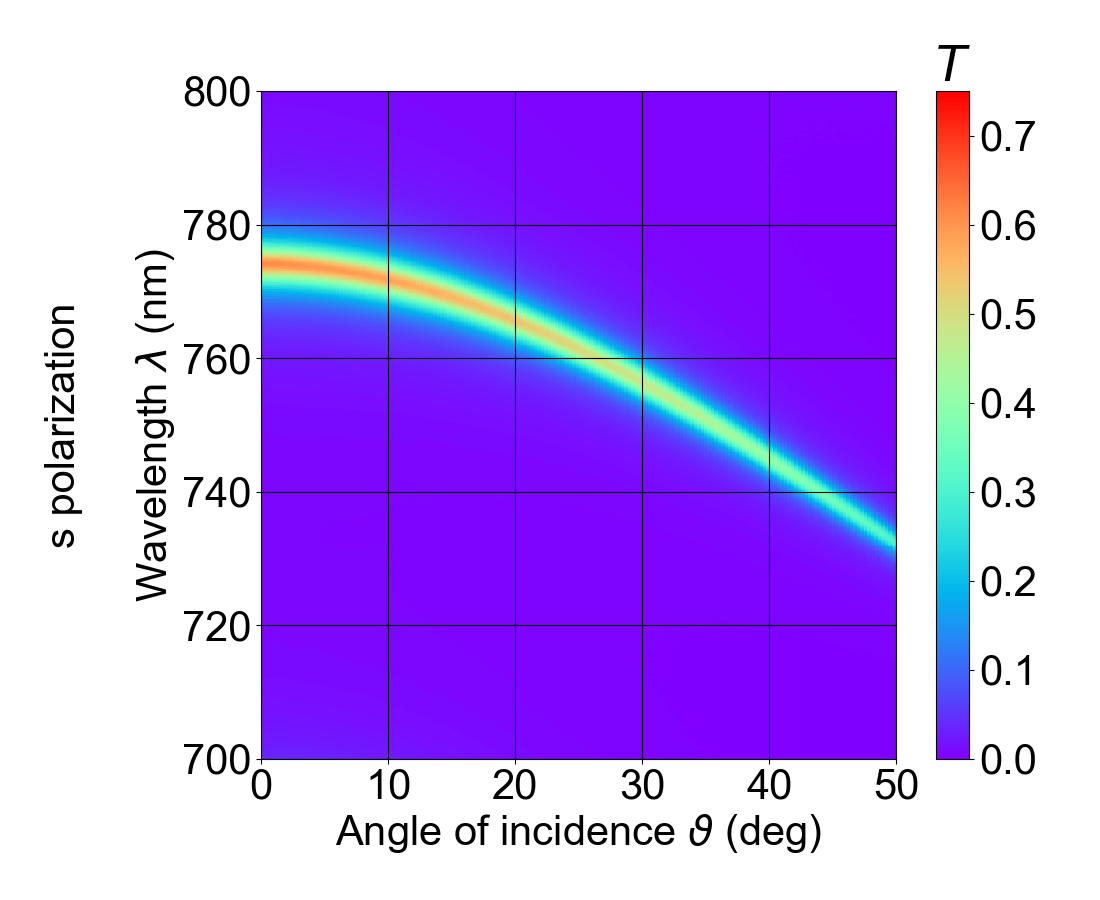}
\includegraphics[width=0.4\linewidth]{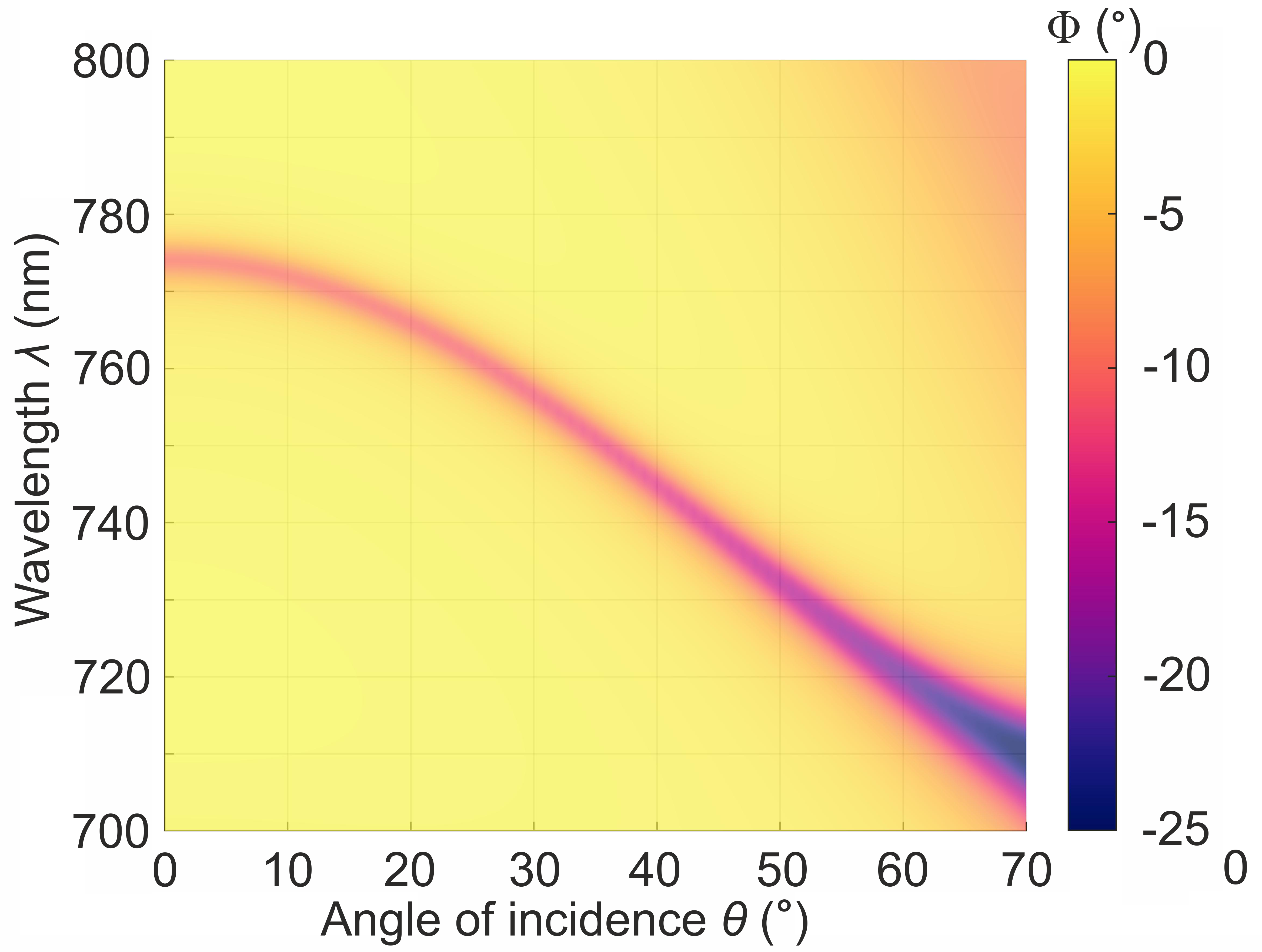} 
\caption{Wavelength vs. incidence angle spectra of (a),(c) experimentally measured transmittance and (b),(d) simulated Faraday rotation $\Phi(+M_z)$ for p- (left pane) and s-polarized (right pane) light.}
\label{fig: SI T and Far}
\end{figure*}

Figure~\ref{fig: SI T and Far} shows the transmittance and Faraday rotation spectra for p- and s-polarized light at different angles of incidence. It is clearly seen that the difference between the two polarization increases with the increase of the angle of incidence. This can be explained by the growing difference between the Fresnel reflection coefficients and consequent difference of the reflectance of the Bragg mirrors surrounding the cavity layer of the MPC. This results in the increase of the Q-factor difference of the p-polarized and s-polarized cavity mode, resulting in the several times different transmittance and Faraday rotation observed for these polarizations (Fig.~\ref{fig: SI T and Far}).

Thus, at the oblique incidence the system is sensitive to the mutual orientation of $\mathbf{k}_\tau$ and $\mathbf{E}$. The more pronounced this difference is, the higher inequality of counter- and clockwise rotations of the intermediate polarisation could be expected. Thus, AFE is expected to increase with the increse of the angle of incidence.

\subsection{Faraday rotation measurements via balanced scheme}

To measure the Faraday rotation $\Phi(M_z)$ one might determine the magneto-optical contribution to the total polarization rotation acquired by the light during the propagation in a magnetic structure. In the most general case, the total polarization rotation $\mathrm{TR}$ has both purely optical and magneto-optical contributions:
\begin{equation}
    \mathrm{TR}=\mathrm{OR}+\Phi(M_z).
\end{equation}

Purely optical polarization rotation might exist in the structure, for example, due to the Fresnel reflection or other magnetization-independent effects. Thus, this contribution should be taken into account and eliminated in the measurements. 

As the Faraday rotation $\Phi$ is usually odd in the magnetization (see Eq.~\eqref{Phi long}), its contribution can be determined in one of the two equivalent ways, as $\Phi(M_z)=\frac{1}{2}(\mathrm{TR}(+M_z)-\mathrm{TR}(-M_z))$ or $\Phi(M_z)=(\mathrm{TR}(+M_z)-\mathrm{TR}(M_z=0))$. For the case where the asymmetric Faraday rotation is expected, one might use the later formula and measure $\mathrm{TR}$ for $+M_z$, $-M_z$ and $M_z=0$ separately.

The measurements of the total polarization rotation $TR$ can be performed in different ways. One of the well-known methods providing high precision is based on a so-called balanced scheme with additional polarizer. If a polarizer is installed after the magnetic structure and aligned at an angle $\alpha$ with respect to the polarization of the incident linearly polarized light, then the intensity of light transmitted through this polarizer equals to 
\begin{equation}
    I_\mathrm{out}=I_0 \cos^2 (\alpha+\mathrm{TR}). \label{Eq Ipol}
\end{equation}

In the balanced scheme, the polarizer is aligned at $\alpha=\pi/4$ angle with respect to the incident polarization of light. Therefore, according to Eq.~\eqref{Eq Ipol} the intensity of light transmitted through the polarizer equals to 
\begin{align}
    I_\mathrm{out}&=I_0 \cos^2 \left(\frac{\pi}{4}+\mathrm{TR}\right) =  \nonumber \\
     &=\frac{I_0}{2} \left(\cos\left(2\left(\frac{\pi}{4}+\mathrm{TR}\right)\right)+1\right)=\nonumber\\
    &=\frac{I_0}{2} \left( 1-\sin\left(2\mathrm{TR}\right)\right)
    \label{Eq I_out}
\end{align}
In order to make it possible to reveal the difference between the values of $\Phi(+M_z)$ and $\Phi(-M_z)$, the following measurements were performed:
\begin{enumerate}
    \item The transmittance $I_0(\lambda)$ of the MPC without additional polarized was measured.
    \item The purely optical rotation 
    \begin{equation}\mathrm{OR}(\lambda)=\frac{1}{2} \mathrm{asin}\left(1- \frac{2 I^{\pi/4}_{M_z=0}(\lambda)}{I_0(\lambda)}\right)
    \end{equation}
    was determined from the measurements of the transmitted intensity $ I^{\pi/4}_{M_z=0}(\lambda)$ in a balanced scheme with additional $\pi/4$-tilted polarizer and in-plane magnetic field applied to the structure to provide $M_z=0$ (see also Sec.~\ref{Sec in-plane})
    \item In accordance to Eq.~\eqref{Eq I_out}, the Faraday rotation for $+M_z$ and $-M_z$ magnetizations was measured as 
    \begin{equation}
        \Phi(\pm M_z)=\frac{1}{2} \mathrm{asin}\left(1- \frac{2 I^{\pi/4}_{\pm M_z}(\lambda)}{I_0(\lambda)}\right)-\mathrm{OR}(\lambda).
    \end{equation}
    The transmitted intensity $ I^{\pi/4}_{\pm M_z}(\lambda)$ was measured separately for $+M_z$ and $-M_z$ magnetization directions in a balanced scheme with additional $\pi/4$-tilted polarizer.
\end{enumerate}

Thus, step-by-step experimental measurements were performed to reveal AFE and to take into account possible changes of the transmitted light intensity and optical rotation due to the purely optical effects.

\bibliography{apssamp}